\documentstyle[aps,epsfig,multicol]{revtex}
\begin{document}
\draft
\title{Momentum noise in a quantum point contact}
\author{A. Tajic, M. Kindermann, and C. W. J. Beenakker}
\address{Instituut-Lorentz, Universiteit Leiden, P.O. Box 9506, 2300 RA Leiden,
The Netherlands}
\date{June 2002}
\maketitle
\begin{abstract}
Ballistic electrons flowing through a constriction can transfer momentum to the lattice and 
excite a vibration of a free-standing conductor. We show (both numerically and analytically) 
that the electromechanical noise power $P$ does not vanish on the plateaus of quantized 
conductance --- in contrast to the current noise. The dependence of $P$ on the constriction 
width can be oscillatory or stepwise, depending on the geometry. The stepwise increase 
amounts to an approximate quantization of momentum noise.
\end{abstract}
\pacs{PACS numbers: 85.85.+j, 73.23.Ad, 73.50.Td, 73.63.Rt}
\begin{multicols}{2}

Not long after the discovery of conductance quantization in a ballistic
constriction \cite{Van Wees-Wharam} it was
predicted that the quantization is noiseless \cite{Lesovik}: 
The time dependent current fluctuations should vanish at
low temperatures on the plateaus of quantized conductance and they
should peak in the transition from one plateau to the next. The
conclusive experimental verification of this prediction followed many
years later \cite{Glattli-Reznikov}, delayed by the
difficulty of eliminating extraneous sources of noise. The notion of
noiseless quantum ballistic transport is now well established
\cite{Blanter-Buttiker}.

The origin of noiseless transport lies in the fact that the eigenvalues
$T_{n}$ of the transmission matrix product $tt^{\dagger}$ take only the
values 0 or 1 on a conductance plateau. The current noise power at zero 
temperature $P_{I}\propto\sum_{n}T_{n}(1-T_{n})$ then vanishes \cite{Buttiker}. 
In other words, current fluctuations require partially filled scattering
channels, which are incompatible with a quantized conductance.

In this paper we point out that the notion of noiseless quantum
ballistic transport does not apply if one considers momentum transfer
instead of charge transfer. Momentum noise created by an electrical
current (socalled electromechanical noise) has been studied in the
tunneling regime \cite{Bocko} and in a diffusive conductor
\cite{Shytov}, but not yet in connection with ballistic transport. Our
analysis is based on a recent scattering matrix representation of the
momentum noise power $P$ \cite{Kindermann}, according to which $P$ depends
not only on the transmission eigenvalues but also on the eigenvectors.
This makes it possible for the electrons to generate noise even in the
absence of partially filled scattering channels.

\begin{figure}[!h]
\begin{center}
\epsfig{file=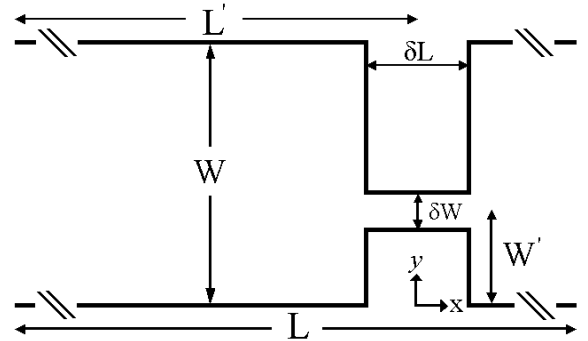,height=4.56cm,width=7.55cm}
\medskip
\caption{Schematic diagram of a two-dimensional channel containing a constriction. 
The current flows in the $x$-direction, the channel is free to vibrate either in the $x$-direction 
(longitudinal vibration) or in the $y$-direction (transverse vibration).}
\label{fig1}
\end{center}
\end{figure}

The geometry is shown schematically in Fig.\ $\ref{fig1}$. We consider a two-dimensional electron
gas channel in the $x$-$y$ plane. The width of the channel in the $y$-direction
is $W$ and the length in the $x$-direction is $L$. The channel contains a narrow constriction of length 
$\delta L \ll L$ and width $\delta W \ll W$ located at a distance $L'$ from the left end.  
(We choose $x=0$ at the middle of the constriction, so that the channel extends from 
$-L'<x<L-L'$.) A voltage $V$ drives a current through the constriction, 
exciting a vibration of the channel. We seek the low-frequency noise power
\begin{equation}
P=2\int_{-\infty}^{\infty}dt\,
\overline{\delta F(0) \delta F(t)}=\lim_{t\rightarrow\infty}\frac{2}{t}\overline{\Delta
{\cal P}(t)^{2}}\label{Pdef}
\end{equation}
of the fluctuating force $\delta F(t) \equiv F(t)-\overline{F}$ that drives the vibration. 
The noise power is proportional to the variance of the momentum $\Delta {\cal P}(t)$ 
transferred by the electrons to the channel in a long time $t$.

Depending on whether the channel is free to vibrate in the $x$ or $y$ direction, we distinguish a 
longitudinal vibration (with amplitude ${\bf u}({\bf r})=u(x)\hat{x}$) and a transverse vibration 
(${\bf u}({\bf r})=u(x)\hat{y}$). If the electron gas is deposited on top of a doubly-clamped beam, 
then the channel should be oriented perpendicular to the beam for a longitudinal vibration and 
parallel to the beam for a transverse vibration.

The wave functions are represented by scattering states. The incident wave has the form $\phi_{n}({\bf
r})=|\hbar k_{n}/m^{\ast}|^{-1/2}\exp(ik_{n}x)\Phi_{n}(y)$, where $m^{\ast}$ is the effective mass, 
$n=1,2,\ldots N$ is the mode
index, $\Phi_{n}$ the transverse wave function, and
$k_{n}=\pm(2m^{\ast}/\hbar^{2})^{1/2}(E_{F}-E_{n})^{1/2}$ the longitudinal wave
vector (at Fermi energy $E_{F}$ larger than the cutoff energy $E_{n}$). We take
$k_{n}$ positive to the left and negative to the right of the constriction.
Incident and outgoing waves are related by the $2N\times 2N$ unitary scattering
matrix

\begin{equation}
S=\left(\begin{array}{cc}
r&t'\\t&r'
\end{array}\right),\label{Sdef}
\end{equation}
containing the $N\times N$ transmission and reflection matrices $t,t',r,r'$. We
assume time reversal symmetry, so that $\Phi_{n}$ is real and $S$ is symmetric.

As derived in Ref.\ \cite{Kindermann}, the noise power $P$ and the mean force 
$\overline F$ for a localized scatterer can be expressed in terms of
the matrix $S$ and a Hermitian matrix ${\cal M}_{nn'}=m^{\ast -1}\sum_{\alpha\beta}\langle
n|p_{\alpha}u_{\alpha\beta}p_{\beta}|n'\rangle$ of expectation values in the
basis of incident modes. The expectation value is taken of the electron
momentum flux $m^{\ast -1}p_{\alpha}p_{\beta}$, weighted by the strain
tensor $u_{\alpha\beta}=\frac{1}{2}(\partial u_{\alpha}/\partial
x_{\beta}+\partial u_{\beta}/\partial x_{\alpha})$. The matrix ${\cal M}$ is
block-diagonal,
\begin{equation}
{\cal M}=\left(\begin{array}{cc}
M_{\rm L}&0\\0&M_{\rm R}
\end{array}\right).\label{Mdef}
\end{equation}
At zero temperature and to first order in the voltage one has, for a two-fold spin 
degeneracy,
\begin{eqnarray}
P&=&\frac{4eV}{h}{\rm Tr}\,\big(rr^{\dagger}M_{\rm
L}^{\ast}t't'^{\dagger}M_{\rm L}^{\ast}
+r'r'^{\dagger}M_{\rm R}^{\ast}tt^{\dagger}M_{\rm
R}^{\ast}\nonumber\\
&&\mbox{}-2tr^{\dagger}M_{\rm L}^{\ast}rt^{\dagger}M_{\rm
R}^{\ast}\big),\label{PSMrelation}\\
\nonumber\\
\overline{F}&=& \frac{2eV}{h}{\rm Tr}\,\big( M_{\rm L}+rr^{\dagger}M_{\rm L}^{\ast}+
tt^{\dagger}M_{\rm R}^{\ast}\big).\label{meanforce}
\end{eqnarray}
In Eq.\ $(\ref{meanforce})$ we have not included the equilibrium contribution to the 
mean force (at $V=0$). Electron-electron interactions (screening) are not accounted for, since we do not expect 
any appreciable charge accumulation in a ballistic system. 

For a longitudinal vibration one has simply $(M_{\rm L})_{nm}=-(M_{\rm R})_{nm}=\delta_{nm}\, \hbar \, |k_{n}|\,u(0)$.
For a transverse vibration the blocks $M_{\rm L},M_{\rm R}$ are not diagonal,
\begin{eqnarray}
(M_{\rm L,R})_{nm}&=&\frac{\hbar(k_{n}+k_{m})}{2i|k_{n}k_{m}|^{1/2}} \int
dy\,\Phi_{n}(y)\Phi'_{m}(y)\nonumber\\
&&\mbox{}\times\int_{\rm L,R} dx\,u'(x)\exp[ix(k_{m}-k_{n})].\label{MLR}
\end{eqnarray}
The integral over $x$ extends over the region $(-L',-\delta L/2)$ to the left of
the constriction for $M_{\rm L}$ and over the region $(\delta L/2,L-L')$ to the
right of the constriction for $M_{\rm R}$. We abbreviate $q_{nm}=k_{m}-k_{n}$.
For $|n-m|$ of order unity one has $q_{nm}$ of order $1/W$, so that the range of $x$
that contributes to the integral is of order $W$. (Contributions from inside the constriction 
are smaller by a factor ${\rm min}\,(\delta W, \delta L)/W$.) Since $W$ is much greater 
than the Fermi wave length $\lambda_{F}$, we are justified in using the asymptotic 
plane wave form of the scattering states to calculate ${\cal M}$. 

We take hard-wall boundary conditions at $y=0,W$,
hence $\Phi_{n}(y)=(2/W)^{1/2}\sin(n\pi y/W)$,
$E_{n}=(\hbar^{2}/2m^{\ast})(n\pi/W)^{2}$, $N=[2W/ \lambda_{F}]$, and
$(k_{n}+k_{m})q_{nm}=(\pi/W)^{2}(n^{2}-m^{2})$. The overlap
$\int_{0}^{W}dy\,\Phi_{n}\Phi'_{m}$ is evaluated straightforwardly, but the
integration over $x$ requires more care. The derivative $u'(x)$ of the mode
profile vanishes at the two clamped ends of the beam, as well as at its center.
We assume that the constriction is off-center, therefore $u'(\pm\delta L/2)\approx u'(0)\neq 0$. 
We write $u'(0)=u_{0}/L$, with $u_{0}$ a 
number of order unity.
Upon partial integration we find, to first order in $W/L$,
\begin{eqnarray}
\int_{\rm L,R} dx\,u'(x)\exp(ixq_{nm})=&\pm&\frac{u_{0}}{iq_{nm}L}\exp(\mp iq_{nm}\delta L/2)\nonumber\\
&+&{\cal O}(W/L)^{2}.\label{uqmn_approx}
\end{eqnarray}
(The upper sign is for region L, the lower sign for region R.) 
We thus arrive for the transverse vibration at the result $M_{L}=-M_{R} \equiv M$, with
\begin{eqnarray}
M_{nm}&=&\frac{2\hbar Wu_{0}}{\pi^{2}L}(\sigma_{nm}-1)
\frac{nm(k_{n}+k_{m})^{2}}
{(n^{2}-m^{2})^{2}|k_{n}k_{m}|^{1/2}}\nonumber\\
&&\mbox{}\times\exp\big[i\,\big(\, |k_{n}|-|k_{m}|\, \big)\,\delta L/2 \big].\label{MLRresult}
\end{eqnarray}
The symbol $\sigma_{nm}=\frac{1}{2}[1+(-1)^{n+m}]$ selects indices of the same parity, so that 
$M_{nm}=0$ if $n$ and $m$ are both even or both odd.

\begin{figure}[!h]
\begin{center}
\epsfig{file=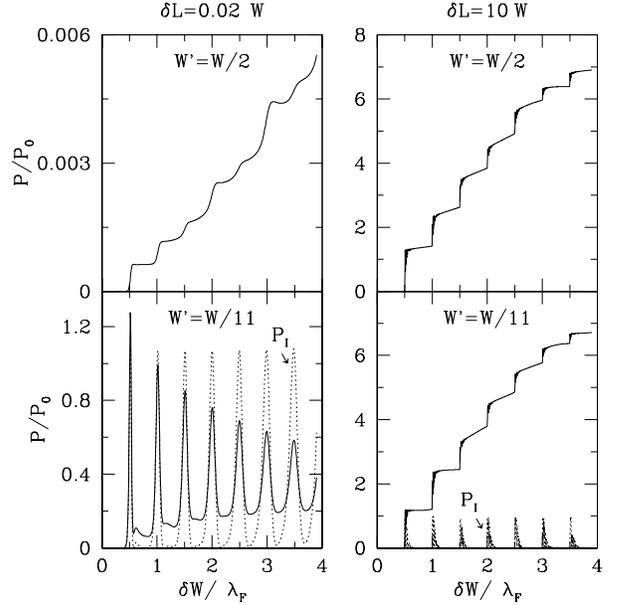,height=8cm,width=8cm}
\medskip
\caption{Solid lines: noise power $P$ for transverse vibration versus width of constriction $\delta W$, at fixed 
width $W=49.9\,\lambda_{F}$ of the wide channel. The left panels are for a short 
constriction with and without axial symmetry. The right panels are the corresponding 
results for a long constriction. The dotted line is the current noise $P_{I}$ in units of 
$e^{3}V/h$ (which is nearly the same with and without axial symmetry).}
\label{fig2}
\end{center}
\end{figure}

\begin{figure}[!h]
\begin{center}
\epsfig{file=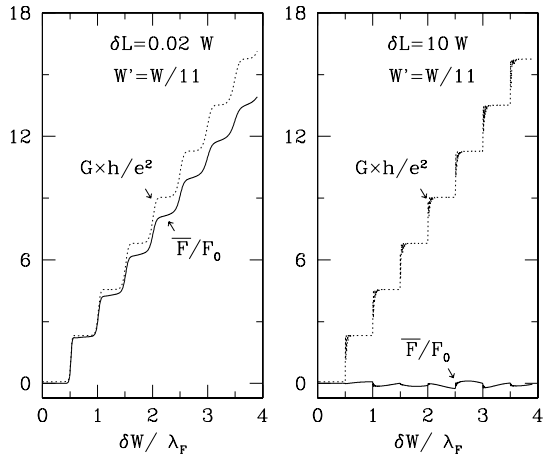,height=8cm,width=8cm}
\vspace*{-1.5 cm}
\caption{Solid lines: average transverse force as a function of constriction width,  
in the absence of axial symmetry (positive values point in the positive $y$-direction 
in the geometry of Fig.\ $\ref{fig1}$, for a current flowing in the positive x-direction). 
The left and right panels are for a short and long 
constriction, respectively. The conductance of the constriction is shown as a dotted line. 
The average transverse force is identically zero for the axially symmetric geometry 
$(W'=W/2)$.}
\label{fig3}
\end{center}
\end{figure}

Our constriction has left-right symmetry, so $r=r'$ and $t=t'$. We contrast the case 
$W'=\frac{1}{2}W$ of axial symmetry with the case $W' \ll \frac{1}{2}W$ of a constriction 
placed highly off-axis. We also contrast the short-constriction case $\delta L \ll W$ 
(point contact geometry) with the long-constriction case $\delta L \gg W$ 
(microbridge geometry). The reflection and transmission matrices are calculated by 
matching wavefunction modes at $x=\pm \delta L/2$, cf.\ Ref.\ \cite{Szafer}.
We first discuss the case of a transverse vibration, which has the richest dependence on the 
geometry.
 
In Fig.\ $\ref{fig2}$ we show the dependence of the transverse noise power $P$ 
[in units of $P_{0}=(4eV/h)(Nu_{0}\hbar/L)^{2}$] on the width $\delta W$ of 
the constriction, at fixed width $W$ of the wide channel. (We choose $W=49.9\,\lambda_{F}$, 
so $N=99$.) The average transverse force $\overline{F}$ is shown in Fig.\ $\ref{fig3}$, normalized by 
$F_{0}=(2eV/h)(Nu_{0}\hbar/L)$. (Note that $\overline{F}=0$ for the axially symmetric case.) 
The conductance $G=(2e^{2}/h){\rm Tr}\,tt^{\dagger}$ and the current noise 
$P_{I}=(4e^{3}V/h){\rm Tr}\,tt^{\dagger}(1-tt^{\dagger})$ are included in these plots for comparison.

The plots show a remarkably varied behavior: For the short constriction without 
axial symmetry the noise power $P$ of the transverse force oscillates as a function 
of the constriction width $\delta W$, in much the same way as the current noise power 
$P_{I}$ oscillates \cite{Lesovik,Buttiker}. However, the minima in $P$ do not go to zero like the 
minima in $P_{I}$, demonstrating non-zero momentum noise on the plateaus of quantized conductance. 
If the short constriction is precisely at the center of the channel, $P$ increases nearly monotonically with $\delta W$. 
For a long constriction $P$ increases nearly monotonically regardless of whether there is 
axial symmetry or not. The increase of the noise power is stepwise, reminiscent 
of the conductance. (The current noise in the long constriction remains oscillatory.)
The mean transverse force behaves similarly to the conductance for the short constriction, but fluctuates 
around zero for the long constriction.

In order to explain the approximate quantization of momentum noise in analytical terms it is convenient to 
decompose the (symmetric) transmission matrix as 
$t_{nm}=\sum_{n'}U_{nn'}U_{mn'}\sqrt{T_{n'}}$, where $U$ is an $N\times N$ unitary 
matrix and $T_{n}\in[0,1]$ is the transmission eigenvalue (eigenvalue of $tt^{\dagger}$). 
Similarly, the reflection matrix is decomposed as 
$r_{nm}=i\sum_{n'}U_{nn'}U_{mn'}\sqrt{1-T_{n'}}$. In this representation 
Eq.\ (\ref{PSMrelation}) takes the form 
\begin{eqnarray}
P&=&\frac{8eV}{h}\sum_{n,m}|X_{nm}|^{2}\bigl[(1-T_{n})T_{m}\nonumber\\
&&\mbox{}+[T_{n}(1-T_{n})T_{m}(1-T_{m})]^{1/2}\bigr],\;\;X=U^{\dagger}M^{\ast}U.\label{PXT}
\end{eqnarray}
The matrix $M$ couples only mode indices of different parity, cf. Eq.\ (\ref{MLRresult}). 
The presence or absence of axial symmetry manifests itself in the matrix $U$, 
which couples only indices of the same parity if $W'=W/2$. In this axially symmetric case 
$X_{nm}=0$ if $n,m$ have the same parity.

In a simple model \cite{Bee97} of a long and narrow ballistic microbridge, $U$ is a random matrix while 
the transmission eigenvalues take on only two values: $T_{n}=1$ for $1\leq n\leq \delta N$ 
and $T_{n}=0$ for $\delta N<n\leq N$. The number $\delta N=[2\delta W/\lambda_{F}]$ is 
the quantized conductance of the constriction (in units of $2e^{2}/h$). Averages of $U$ 
over the unitary group introduce Kronecker delta's (cf.\ App.\ B in Ref.\ \cite{Bee97}). 
We need the average $\Omega_{pp'qq'nm}\equiv\langle U^{\ast}_{pn}
U^{\ast}_{qm}U^{\vphantom{\ast}}_{p'm} U^{\vphantom{\ast}}_{q'n}\rangle$, given by
\begin{equation}
\Omega_{pp'qq'nm}=\frac{1}{N^{2}-1} \left(\delta_{pq'}\delta_{qp'} -
\frac{1}{N}\delta_{pp'}\delta_{qq'}\right)\;\;{\rm if}\;\;n\neq m,\label{Uaverage1}
\end{equation}
in the case of broken axial symmetry and
\begin{equation}
\Omega_{pp'qq'nm}=\frac{4}{N^{2}-\sigma_{N1}} \delta_{pq'}\delta_{qp'} 
\sigma_{pn}\sigma_{qm}\;\;{\rm if}\;\;\sigma_{nm}=0,\label{Uaverage2}
\end{equation}
in the axially symmetric case.

Substituting these values of $T_{n}$ into Eq.\ (\ref{PXT}) and averaging over $U$ with 
the help of Eqs.\ (\ref{Uaverage1}) and (\ref{Uaverage2}), we find
\begin{eqnarray}
P&=&\frac{8eV}{h}\sum_{n=\delta N+1}^{N}\sum_{m=1}^{\delta N}\sum_{p,p',q,q'=1}^{N}
\Omega_{pp'qq'nm} M^{\ast}_{pp'}M^{\vphantom{\ast}}_{q'q}\nonumber\\
&=&\frac{8eV}{h}\frac{\delta N}{N}\left(1-\frac{\delta N}{N}\right){\rm Tr}\,M^{2}\nonumber\\
&=&\frac{\pi^{2}}{9}\delta N P_{0},\;\;N\gg \delta N,\label{PQPC}
\end{eqnarray}
regardless of whether axial symmetry is present or not. We thus 
obtain a stepwise increase of $P$ as a function of $\delta W$ with step height 
$\Delta P=(\pi^{2}/9)P_{0}$. The numerically obtained step height in Fig.\ $\ref{fig2}$ agrees within 10\% 
with the analytical estimate for the first step. For subsequent steps the agreement becomes worse, presumably 
because the approximation of a uniform distribution of $U$ breaks down as $\delta W$ increases. 
We can also calculate the mean transverse force in the same way, starting from 
Eq.\ (\ref{meanforce}), and find $\overline{F}\propto{\rm Tr}\,M=0$, in accordance 
with the numerical result that $\overline{F}\ll F_{0}$ for a long constriction.

In the short-constriction case $\delta L\ll W$ we may not treat $U$ as uniformly distributed in the unitary group, 
and this has prevented us from finding a simple analytical representation of the numerical data.

\begin{figure}[!h]
\begin{center}
\epsfig{file=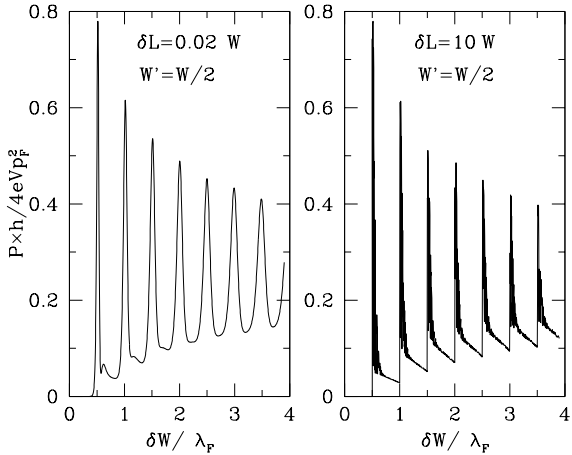,height=8cm,width=8cm}
\vspace*{-1.5 cm}
\caption{Noise power for longitudinal vibration of a short constriction (left panel) and 
a long constriction (right panel).  These plots are for $W'=W/2$, but there is no noticeable 
dependence on the ratio $W'/W$. The mean longitudinal force (not shown) decreases stepwise as 
a function of $\delta W$ in both the short and long constriction. }
\label{fig4}
\end{center}
\end{figure}

In Fig.\ $\ref{fig4}$ we show the noise power for longitudinal vibration. It does not depend on the presence or 
absence of axial symmetry and is also rather insensitive to the length of the constriction. The order of magnitude 
of the longitudinal noise power is $(4eV/h)p_{F}^{2}$, with $p_{F}=\hbar k_{F}$ the Fermi momentum. This is larger 
than the typical transverse noise power $P_{0}$ by a factor of order $(k_{F}L/N)^{2}\simeq (L/W)^{2}$. 
Inserting parameters $V=1\, {\rm mV},\, k_{F}=10^{8}\, {\rm m}^{-1}$, typical for a two-dimensional electron gas, 
one estimates $(4eV/h)p_{F}^{2}\simeq 10^{-40}\, {\rm N}^{2}/{\rm Hz}$. This is below the force sensitivity of present 
day nanomechanical oscillators, but is hoped to be reached in future generations of these devices \cite{Roukes}.

In summary, we have demonstrated that the momentum noise of ballistic electrons does not vanish on the plateaus of 
quantized conductance. Conductance quantization requires absence of backscattering in the constriction, but it 
does not preclude inter-mode scattering. Momentum noise makes this inter-mode scattering visible in a way that 
current noise can not. The dependence of the momentum noise on the constriction width was found to be remarkably 
varied, ranging from oscillatory to stepwise, depending on the direction of the vibration (longitudinal or transverse 
to the constriction), the presence or absence of axial symmetry, and the length of the constriction. The stepwise 
increase amounts to a quantum of momentum noise that might be observable with an ultrasensitive oscillator. 

This research was supported by the Dutch Science Foundation NWO/FOM.

\end{multicols}
\end{document}